# Einstein's 1912-1913 struggles with Gravitation Theory: Importance of Static Gravitational Fields Theory

## Galina Weinstein[*]

*January 31, 2012*

In December 1911, Max Abraham published a paper on gravitation at the basis of which was Albert Einstein's 1911 June conclusion about a relationship between the velocity of light and the gravitational potential. In February 1912, Einstein published his work on static gravitational fields, which was based on his 1911 June theory. In March 1912, Einstein corrected his paper, but Abraham claimed that Einstein borrowed his equations; however, it was actually Abraham who needed Einstein's ideas and not the other way round. Einstein thought that Abraham converted to his theory of static fields while Abraham presumed exactly the opposite. Einstein then moved to Zurich and switched to new mathematical tools. He examined various candidates for generally covariant field equations, and already considered the field equations of his general theory of relativity about three years before he published them in November 1915. However, he discarded these equations only to return to them more than three years later. Einstein's 1912 theory of static fields finally led him to reject the generally covariant field equations and to develop limited generally covariant field equations.

## 1. Static Fields and Polemic with Max Abraham

In June 1911 Einstein's published in the *Annalen der Physik* his paper, "Uber den Einfluβ der Schwerkraft auf die Ausbreitung des Lichtes" ("On the Influence of Gravitation on the Propagation of Light"). An important conclusion of this paper is that the velocity of light in a gravitational field is a function of the place: if it is $c_0$ at the origin of the coordinates, then at a place with a gravitational potential $\Phi$ it is given by the equation:

$c = c_0(1 + \Phi/c^2).$[1]

The above equation signifies that there exists a relationship between the velocity of light and the gravitational potential; the latter influences the first.

Accordingly, beginning in 1912 Einstein claimed that the velocity of light determined the field and thus he offered a theory of static fields which violated his own light postulate from the special theory of relativity,[2] and as a consequence "this result excludes the general validity of the Lorentz transformation; it must not deter us from further pursuing the chosen path".[3]

---

[*] This paper was written while I was a visiting scholar in the Center for Einstein Studies, Boston University.

Einstein's 1911 *Annalen* paper drew the attention of other scientists to develop their own gravitation theory. In December 1911, a short time after the publication of Einstein's 1911 *Annalen* paper, Max Abraham from Göttingen submitted a paper to an Italian journal and translated it to German for the *Physikalische Zeitschrift*, "Zur Theorie der Gravitation" ("On the Theory of Gravitation"). Abraham formulated his theory in terms of Hermann Minkowski's four-dimensional space-time formalism and Einstein's above 1911 relation between the variable velocity of light and the gravitational potential. [4]

Since Minkowski took the constancy of the speed of light to be one of his fundamental principles, Abraham Pais wrote that Abraham tried the impossible: to incorporate Einstein's idea of a non-constant light velocity into the special theory of relativity.[5] Eat one's cake and have it too…

In February 1912 Abraham published a Berichtigung (correction) to his paper:[6]

In lines 16 and 17 of my note 'On the Theory of Gravitation" an oversight has to be corrected which was brought to my attention by a friendly note from Mr. A. Einstein. Hence one should read there 'we consider *dx, dy, dz*, and $du = idl = icdt$ as components of a displacement *ds* in four dimensional space".

Abraham then got the following idea as a result of Einstein's correction: [7]

"Hence:

$$ds^2 = dx^2 + dy^2 + dz^2 - cdt^2$$

is the square of the four-dimensional line element, where the speed of light *c* is determined by eq (6)".

Jürgen Renn explained that, "Abraham had effectively introduced […] the general four-dimensional line element involving a variable metric tensor. However, for the time being Abraham's expression remained an isolated mathematical formula without context and physical meaning which, at this point, was indeed neither provided by Abraham's nor by Einstein's physical understanding of gravitation."[8]

From Abraham's above Berichtigung we can infer that Einstein read Abraham's paper of 1912 on the theory of gravitation, "Zur Theorie der Gravitation"; he corrected it, and then responded to it by a theory of his own. In February 1912, simultaneously to Abraham's correction, Einstein submitted his paper, "the Speed of Light and the Statics of the Gravitational Fields".[9]

Einstein explained that after he had published his 1911 *Annalen* paper,[10]

"Abraham has created a theory of gravitation which renders information and draws conclusions from my first paper as a special case. But we shall see in what follows that Abraham's system of equations cannot be reconciled with the equivalence hypothesis, and that his conception of time and space already does not hold up from a purely formal, mathematical point of view".

On February 20, 1912, Einstein wrote Ludwig Hopf "Abraham's theory is completely wrong" And Einstein anticipated differences between him and Abraham.[11] Einstein attacked Abraham's theory because of what he considered as the incompatibility between Abraham's simultaneous implementation of both a variable speed of light and Minkowski's formalism; and because Abraham's theory could not be reconciled with a theory based on the equivalence principle.

On February 24, 1912, Einstein wrote the editor of the *Annalen*, Wilhelm Wien,[12]

"I am sending you here a paper for the Annalen. Viewed on it many drops of sweat, but I now have complete confidence in the matter. Abraham's theory of gravitation is completely unacceptable. How could anybody be so lucky to guess the correct equations without any effort! Now I am searching for the dynamics of gravitation. It will proceed but not so quickly!"

On February 26, 1912 Einstein's paper was received. The "many drops of sweat" were probably left here: Einstein used his technique from his 1911 *Annalen* paper of comparing an accelerated system $K(x, y, z, t)$ and a system of constant gravitational potential $\Sigma(\xi, \eta, \zeta, \tau)$. The $\xi$–axis coincides with the $x$–axis and the $\eta$–axis is parallel to the $z$–axis. Einstein found that the following equations hold good in system $K$,

$\xi = x + act^2/2, \eta = y, \zeta = z, \tau = ct,$

where,

$c = c_0 + ax.$

$a/c_0 =$ acceleration of the origin of $K$ with respect to $\Sigma$.[13]

According to the above equation, the equation,

$\Delta c = 0$

is satisfied in $K$.

This is a linear equation that corresponds to Poisson's equation in Newtonian theory:

$\Delta c = \kappa c \rho,$

where $\kappa$ denotes the universal gravitational constant, and $\rho$ the matter density ($\Delta$ is the Laplacian operator).

Einstein now wanted to obtain the law of motion of a material point in a static gravitational field.[14] Guided by the equivalence principle, the same equation at which he had arrived for $K$ should hold good for $\Sigma$. Einstein thus obtained for $\Sigma$, or for a motion of a material point in a static gravitational field the same form of equation,

$$\Delta \Theta = \kappa c^2 \rho,$$

where, $K = \kappa c^2$, a gravitational constant, and $\kappa = K/c^2$ is the universal gravitational constant as before.[15]

Einstein derived equations that include the energy of the material point in a stationary gravitational field. According to special relativity this is related to the mass of the material point. The kinetic energy cannot be distinguished from the gravitational energy, and it depends on the mass, the velocity of the material point, and on the velocity of light. Thus it depends on the gravitational potential. The energy of a point at rest in the gravitational field is $mc$.[16]

In section §3 Einstein reformulated his findings from his 1911 *Annalen* paper considering the time, "If we measure time in $S_1$ [lower gravitational potential] with a clock $U$, *we must measure the time in $S_2$* [higher gravitational potential] *with a clock that goes $1 + \Phi/c^2$ slower than the clock $U$ if you compare it with the clock $U$ in the same place*".[17] We can summarize this insight as "gravity bends time".[18]

And Einstein also realized that since mass and energy are equivalent (different forms of the same thing), "The energy $E_1$ arriving at $S_1$ is greater than the energy $E_2$, measured by the same means, which was emitted in $S_2$, and that being the potential energy of the mass $E_2/c^2$ in the gravitational field";[19] and "The increase in *gravitational* mass is equal to $E_2/c^2$, and therefore is equal to the increase in *inertial* mass resulting from the theory of relativity".[20] Related to this finding is another finding, the above finding of Einstein that the kinetic energy cannot be distinguished from the gravitational energy, and it depends on the mass.

On February 29, 1912, Einstein wrote his friend Heinrich Zangger, "During my research on gravitation I discovered that Abraham's theory (Phys. Zeitschr. No. 1) is completely untenable".[21]

On March 11, 1912 Einstein wrote Wien again and asked him to return him back and not publish the manuscript (submitted on February 24 1912). Einstein very likely already realized that the linear field equation $\Delta c = kc\rho$ should be replaced by a non-linear field equation. And he wrote Wien that he discovered that not everything in the paper was proper.[22] He finally decided to let publish this paper, and to immediately submit another correcting paper. He informed Wien that he was now almost finished with the static field, and was going to discover the laws of *the dynamic field*. He did not forget Abraham, and he wrote Wein, "So einfach, wie Abraham meint, ist die Angelegenheit aber nicht" (the matter is not as easy as Abraham thinks).[23]

Sometime between February 1912 and March 1912, Einstein arrived at a small breakthrough in the theory of the static gravitational field: a non-linear field equation and the realization that the theory is a dynamical theory. Einstein realized that energy would exert gravitational influence just as mass would. Energy contained within the gravitational field itself would also exert gravitational influence. Any change in the strength of the gravitational field would produce an extra variation as the change in the energy within the field fed back into the system as a whole. In other words, Einstein's February 1912 linear equations were not consistent with the principle of action and reaction and the principles of energy and momentum.

On March 20, 1912, Einstein submitted a correcting paper to the *Annalen*,[24] and six days later on March 26, he wrote to his eternal sounding board, his best friend Michele Besso, about his new finding. He started his letter to Besso by telling him, "In recent times, I have been working like a mad on the problem of gravitation. Finally, I have come to finish with the statics. I do not know anything about the dynamic field that will follow only now. I write to you a few results".[25]

Einstein defined a "Fesrsetzung" according to which "While the vel. of light $c$ depends on the location, however it does not depend on the direction". When $c$ = const., we arrive at the ordinary theory of relativity. He explained to Besso that his new theory of gravitation perfectly corresponded with special theory (equivalence of mass-energy).[26]

Einstein included the basic electromagnetic equations in his theory of the static gravitational field. In this case the source of the gravitational field is the density of ponderable matter augmented with locally measured energy density.[27] In his letter to Besso, and in his March 20 paper on static gravitational fields, in section §3, "Great thermal and Gravitational Field", Einstein first presented the linear equation for the static gravitational field from his first paper, but with a little change,

$\Delta c = 0$ for space free of matter, and when there is matter, $\Delta c = \kappa c\sigma$,

where, $\sigma$ denotes the mass density and the energy density. [28]

Then he discovered that the principle of equivalence is valid only locally, [29]

"this important first step is therefore difficult because I am departing for it from the foundations of the unconditional principle of equivalence. It seems that the latter holds only for infinitely small fields. Our derivation of the equations of motion of the material point and of the electromagnetic equations is therefore not illusory, since [the above equations] apply only to infinitely small space".

For the more general case Einstein arrived at a non-linear equation for the static gravitational field:

$c\Delta c - 1/2(\text{grad}c)^2 = \kappa c^2\sigma$.

The second term of the left hand side of the equation is the energy density of the gravitational field multiplied by $c$.[30]

Einstein concluded, [31]

"If every energy density ($\sigma c$) generated a (negative) divergence of the lines of force of gravitation, then this must also hold for the energy density of gravitation itself. We write [the above equation] in the form,

$$\Delta c = \kappa\{c\sigma + 1/2\kappa\,\text{grad}^2\,c/c\}.$$

Then one recognizes immediately that the second term in the brackets is the energy density of the Gravitational field".

Einstein did not forget Abraham; he wrote to Besso in the same letter from March 26, "Abraham's theory was created of the top of his head, i.e., from mere mathematical beauty considerations, torn off and completely untenable".[32] That was almost Abraham's opinion of Einstein's theory, except for the mathematical beauty.

On May 17, 1912, Einstein wrote Wien, and he also wrote Zangger three days later, that Abraham wrote him that he no longer adhered to his own equations, but he converted to Einstein's theory.[33] On June 5, 1912, Einstein again wrote Zangger that he was engaged in an amusing polemic with Abraham but "Abraham had accepted my main new results concerning gravitation".[34]

However, Abraham understood that Einstein converted to his theory. According to Abraham's understanding, Einstein corrected his February theory because he borrowed some equations from him. In his June 1912 reply to Einstein, Abraham said that it would be careless to reject Einstein's results, some of which (expression for the energy density) were precisely similar to those found in Abraham's theory; results that Einstein independently formulated by the equivalence hypothesis. [35]

Abraham started his attack by saying, "Already a year ago, A. Einstein has given up the essential postulate of the constancy of the speed of light by accepting the effect of the gravitational potential on the speed of light, in his earlier theory;[36] in a recently published work [February 1912][37] the requirement of the invariance of the equations of motion under Lorentz's transformations also falls, and this gives the death blow to the theory of relativity. Those who repeatedly went after the sirens songs of this theory should be warned that they might be pleased to note that even its author himself is now convinced by its inconsistency".[38]

Abraham's final criticism was of Einstein's March 20 paper.[39] Abraham did not like Einstein's way of arriving at his results, even after the March correction. He did not like Einstein's use of the "Equivalence Hypothesis", and the correspondence between reference systems. It appeared to Abraham as a fluctuating basis, because Einstein did not yet adopt the Space-Time formalism of relativity (that is, he did not formulate his theory on the basis of Minkowski's space-time formalism). [40]

On July 4, 1912, Einstein replied to Abraham and explained to him that the theory of relativity is correct to the extent to which its two underlying principles are accepted.[41] "As it stands now", asks Einstein in his reply to Abraham, "what is the limit of the two principles" of the theory of relativity? Einstein thinks that the principle of the constancy of the velocity of light can be maintained only insofar as one restricts oneself to spatio-temporal regions of constant gravitational potential. "This is, in my opinion, not the limit of validity of the principle of relativity, but is that of the constancy of the velocity of light, and thus of our current theory of relativity".[42]

Einstein explained to Abraham, "This situation, in my opinion, by no means implies the failure of the principle of relativity, just as the discovery and correct interpretation of Brownian motion did not lead to the consideration of Thermodynamics and Hydrodynamics as heresies". The present theory of relativity would always retain its significance as the simplest theory for the important limiting case of spatio-temporal events in a constant gravitational potential.[43]

Einstein described his equivalence principle of 1912, it could only apply consistently to infinitely small spaces, and Einstein added that he knew that it did not supply a satisfactory basis, "But therein I do not see any reason for also rejecting the equivalence principle because it applies to the infinitely small, no one can deny that this principle is a natural extrapolation of the most general experimental propositions of physics".[44]

Finally, Einstein answered Abraham's plagiarism blames,[45]

"[…] this result contradicts the fundamental equations of Abraham's theory […] Abraham further claims that I had used his expressions for the energy density and the stresses in a gravitational field. This is not true"; and Einstein briefly demonstrated why his expression actually contradicted Abraham's premises. According to Einstein's theory one obtains a certain expression for the energy density in a static gravitational field, while according to Abraham's theory the expression for the energy density is completely different.[46]

On July 25, 1912, Abraham replied, "I cannot understand what sense has Hr. Einstein's reply, if he gives up the 'equivalence hypothesis'".[47] Abraham then showed that his expression for the energy density in a static field, which followed from his 1912 theory of gravitation, exactly coincided with Einstein's expression for the energy density in the field.[48]

On August 16, 1912, Einstein wrote to Ludwig Hopf, "Recently, Abraham – as you may have seen – slaughtered me along with the theory of relativity in two massive attacks, and wrote down (phys. Zeitschr.) the only correct theory of gravitation (under the 'nostrification' of my results) – a stately steed, that lacks three legs! He noted that the knowledge of the mass of energy comes from – Robert Mayer".[49]

Marcel Grossmann, Einstein's loyal friend from school, the Zürich Polytechnic, now called Eidgenössische Technische Hochschule (ETH), was the dean of the department of mathematics and physics of the institute. He assisted Einstein and persuaded his colleagues to offer Einstein a professorship in the ETH. In winter 1911-1912 the decision was made, and Einstein left Prague, after he stayed there less than two years. In July 1912 he returned to Zürich, the place he loved so much, to his youth school, there he stayed a professor until he left to Berlin in the spring of 1914.

In Zürich Einstein decided to be publicly silent. He did not want to continue the "amusing polemic" with Abraham. Einstein thus sent a very short note under the title "Comment on Abraham's Preceding Discussion 'Once Again, Relativity and Gravitation'", to the *Annalen*. He wrote that since each of them has presented his own stand point, he did not think that it was relevant to respond to Abraham's note again. Einstein asked the reader not to interpret his silence as an agreement.[50]

## 2 The Zurich Notebook

Recall that Abraham criticized Einstein for not yet adopting Minkowski's space-time formalism.[51] Einstein understood that Minkowski's formalism was crucial for the further development of his theory of gravitation; however he applied Minkowski's formalism once he had recognized that the gravitational field should be described by the metric tensor field, a mathematical object of ten independent components.

Starting in August 1912, Einstein went through a long odyssey in the search after the correct form of the field equations of his new theory. Einstein began to collaborate with his school-mate the mathematician Grossmann on the theory of gravitation, and Grossmann gradually gave Einstein more and more mathematical tools. The first trial of Einstein's efforts appear to be documented in a blue bound notebook – known as the "Zurich Notebook" – comprised of 96 pages, all written in Einstein's hand. The notebook was found within Einstein's papers after his death.[52]

The back cover of the Notebook bears the title "Relativität" in Einstein's hand (probably an indication that he began his notes at the end). Two pieces of paper were probably taped later to the front of the notebook by Einstein's secretary, Helen Dukas. The subject matter of the calculations in the notebook includes statistical physics, thermodynamics, the basic principles of the four-dimensional representation of electrodynamics, and the major part of the notebook is gravitation.

The calculations that Einstein had done in the final pages of the Notebook indicate continues path towards the paper of 1913 written with Grossmann, "Entwurf einer verallgemeinerten Relativitätstheorie und einer Theorie der Gravitation" ("Outline of a Generalized Theory of Relativity and of a Theory of Gravitation") – the "Entwurf" paper.[53]

Most of the calculations in the Zurich Notebook are extremely sketchy, display a lot of false starts, and of course come with no explanatory text. The gravitation part of the Notebook consists of two parts beginning from the back and from the front, respectively. Einstein seemed to have started from the back of the notebook and he also filled the front of the notebook. Hence the scholars had numbered the pages according to their understanding of Einstein's proceeding in the notebook. The two parts of the notebook meet on pages 43L$_A$ and 43 L$_B$, which are upside down relative to one another. The letter "L" signifies "left-hand" pages and the letter "R" – "right-hand" pages. Part A was written beginning from the back of the notebook and consists of pages 32L – 43L. Part B comprises the pages beginning from the front of the notebook, pages 01L – 31L.

## 2.1 The Back Side of the Notebook

### Minkowski's Line Element

The back side of the notebook begins with doodles and then follows a page that deals with Minkowski. In this page, numbered 32R, Einstein wrote the four space-time coordinates *x, y, z, ict*, and $x_1, x_2, x_3, x_4$.[54] On page 32R Einstein wrote "Diesen Skalar…weider Vierervektor". Einstein considered Minkowski's four-vectors. He made calculations on "Elektrodynamik" and relativity, calculated and erased for another ten or more pages,[55] and then on page 39L wrote the general four-dimensional line element:

$$ds^2 = \Sigma G_{\lambda,\mu} dx_\lambda dx_\mu.[56]$$

This is very likely the first time Einstein has written down this expression, because the coefficients $G_{\mu\nu}$ of the "metric tensor" are written with an upper case *G*. After page 40R Einstein switched to the lower case *g*.[57]

### Gravitational equations and Static Gravitation Theory

In the lower half of page 39L Einstein wanted to find the gravitational equations, how $g_{\mu\nu}$, the components of the gravitational field or the metric tensor, are generated by source masses. There Einstein considers a "Spezialfall für G$_{\lambda,\mu}$" – a special case excepting $G_{44} = c^2$, the case of a coordinate system in which the metric is flat. If $c^2$ is **constant** then this metric represents the Minkowski space-time of **special relativity**. If $c^2$ is a **function of the spatial coordinates**, $c^2 = c^2(x_1, x_2, x_3)$, it represents the metrical generalization of Einstein's 1912 theory of **static gravitational theory**.

At the bottom of the page, Einstein tried to apply the gravitational field equation from his 1912 theory of static gravitational fields. He expected that in the special case of a static field, and with an appropriate choice of coordinates, the gravitational field equations would reduce to his March 20, 1912 nonlinear static field equation:

$$c\Delta c - 1/2(\text{grad} c)^2 = \kappa c^2 \sigma.[58]$$

Einstein thought the number of gravitational potentials would reduce from ten to a single potential, but he confronted the more difficult problem, to reverse this reduction. He wanted to put this equation into malleable form which allowed it to be one component of the ten component metric tensor field equation. Einstein then rewrote the left hand-side of the above equation in terms of the components of the metric tensor of $G_{44} = c^2$ from the "Spezialfall".[59]

On the facing page, 39R Einstein continued to examine this approximation but he failed to find covariant formulation in this direction. He then wrote the following operation from Minkowski's four-dimensional space-time formalism: Div $\Gamma = 0$, where $\Gamma$ denoted the metric tensor, and he asked: "Ist dies invariant?" and he continued to calculate and arrived at the conclusion that it was not invariant.[60]

**A Geodesic Curve and a Mirror Page**

Einstein continued with more calculations for a few more pages until page 41R.[61] On page 41R he showed that – using Newtonian equation of motion of a particle which is not subject to external forces but is constrained to move on a curved surface – the trajectory traced by the particle in the surface is a geodesic, a curve of shortest distance.[62]

Page 43L$_A$ is almost an exact mirror image of page 43 L$_B$. Until page 43L$_A$ Einstein was calculating from one side of the notebook. And until page 43 L$_B$ he was calculating from the back of the notebook upside down, and then the two pages meet in the middle of the notebook. The notebook now proceeds to pages full of calculations of heat theory and thermal physics.[63]

**2.2 The Front Side of the Notebook**

**Gravitational Field Affects Matter and Matter Generates Gravitational Fields**

Einstein came back to gravitation after a few pages on page 05R, and he started to calculate from the front of the notebook. He wrote the title: "Gravitation" and by now Grossmann had probably given Einstein more mathematical tools, because the mathematical style has changed abruptly as compared to the "beginning" of the notebook. [64] Einstein was seeking to find the stress-energy tensor for the gravitational field associated with a gravitation tensor he had already introduced. [65]

At the top of page 05R Einstein started with the equations of motion of a point particle in a metric field from an action principle. The action integral is the proper length of the particle's worldline. The equations of motion of the point particle are written in the form of the Euler-Lagrange equations. Einstein then applied this to a cloud of pressureless dust particles in the presence of a gravitational field: he generalized these results to expressions for the momentum density and the force density in the case of the cloud of dust, and identified the expression for momentum density as part of the stress-energy tensor for pressureless dust.

Subsequently Einstein inserted this stress-energy tensor and a similar expression for the density of force acting on the cloud of dust into his initial equations of motion. On p. 05R he thus arrived at a candidate for the law of energy-momentum conservation in the presence of a gravitational field; and he wrote an equation that expresses the vanishing of the covariant four-divergence of the stress-energy tensor $T_{\mu\nu}$ (The equation expresses the requirement that the stress-energy tensor has zero divergence). Thus conservation of energy-momentum of the matter field is satisfied. The equation is generally covariant.[66]

In the next pages Einstein had become more sophisticated mathematically, but still **was not acquainted with the Riemann tensor**. What was very noticeable was that Einstein used methods due to Eugenio Beltrami. These calculations could either be part of the field equations or play a role in their construction, but they did not yet lead to any promising candidates for the left-hand side of the field equations; they led to several important techniques, results and ideas that Einstein was able to put to good use once he learned about the Riemann tensor. He was now going to investigate the covariance properties of various expressions in order to find generally covariant field equations. Einstein tried to form invariant quantities from the metric tensor, and he tried to extract field equations in this way. [67]

### The "Grossmann Tensor": Riemann Tensor

On page 14L Einstein systematically started to explore the Riemann tensor. At the top of page 14L Einstein wrote on the left: "Grossmann Tensor vierter Mannigfaltigkeit" ("Grosmann's tensor four-manifold"), and next to it on the right he wrote the fully covariant form of the Riemann tensor.[68] Therefore, Grossman very likely conveyed to Einstein the knowledge about the Riemann tensor.

This signified a new stage in Einstein's search for gravitational field equations. In the course of this exploration he considered candidate field equations based on the Ricci tensor which he would come back to only on November 4, 1915. He explored the Ricci tensor for a few pages, and on page 26L he rejected his results from the few pages starting from p. 14L. He finally ended with limited generally covariant field equations, the "Entwurf" field equations.[69]

Pais says that the **transition to Riemannian geometry** must have taken place during **the week prior to August 16, 1912** – and Pais dates this according to the letter that Einstein had written to Ludwig Hopf, and he adds "these conclusions are in harmony with my own recollections of a discussion with Einstein […]".[70] On August 16, 1912, Einstein wrote Hopf the following: "With gravitation it is going brilliant. If I am not deceived, I have now found the most general equations".[71] "The most general equations" are the ones that Einstein wrote on **page 14L** of the Zurich Notebook.

Thus the 14L page of the Zurich Notebook – in which the Riemann tensor made its first appearance – could very likely be dated to **the week prior to August 16, 1912 and sometime around this date**. The pages preceding the 14L page – in which Einstein was searching for candidates – could presumably be dated to the beginning of August 1912.

### Newtonian Limit – "Too Complicated"

Einstein contracted the fourth rank Riemann tensor to form the covariant form of the second rank Ricci tensor. At the bottom of the page Einstein checked whether the Ricci tensor reduces in **the weak field approximation to the Newtonian limit**. Einstein wrote this in the form of an equation in which three of its four second order derivative terms vanish in the weak field approximation. He then wrote below the expression: "Sollte verschwinden" ("should vanish").[72]

The problem – how to cause the Newtonian limit to appear – was going to bring Einstein straight to pages 26L and 26R and the limited covariant field equations. Meanwhile Einstein tried to manipulate the Riemann tensor.[73] At some stage he was frustrated with the mathematics that Grossmann had brought to him and he wrote on page 17L, "zu umstaendlich" (too complicated).[74]

### Newtonian Limit – the Harmonic Coordinate Condition

Generally covariant equations hold in all coordinate systems, whereas the equations of Newtonian gravitation theory do not. In the process of recovering the Poisson equation of Newtonian theory for weak static fields from a generally covariant theory, it is thus necessary to restrict the set of coordinate systems under consideration. This is achieved through coordinate conditions that must also be satisfied by the final solution. On page 19L Einstein then tried to **recover the Newtonian limit** from his generally covariant field equations **with the harmonic coordinate condition** (used to eliminate unwanted second order derivative terms from the Ricci tensor).[75]

Calculations under the heading, "Nochmalige Berechnung des Ebenentensors" ("Repeated Calculations of Surface-tensors"): the Ricci tensor from page 14L again, and the calculation "bleibt stehen" (terminates) at the top of the page. The field "Gleichungen" that Einstein got on page 19L in first order approximation were written again on page 19R:

$$\sum \gamma_{\kappa\kappa} \frac{\partial^2 g_{im}}{\partial x^2_{\kappa}} = -\kappa \rho_0 \frac{dx_i}{ds} \frac{dx_m}{ds} g_{ii} g_{mm}.$$

The left hand side is the core-operator term of the reduced Ricci tensor. The right hand side gives the covariant stress-energy tensor for pressureless dust multiplied by the gravitational constant $\kappa$.

Einstein constructed the field equations out of the Ricci tensor that satisfy the case of first order, weak field approximation; Einstein thus recovered the equations of the Newtonian limit. [76]

**Energy-Momentum conservation – the "Hertz Condition"**

However, as Einstein was examining the generally-covariant Ricci tensor expression to determine whether a physically acceptable field equation could be extracted from it, he assumed that he needed an additional condition, **not just to recover the Poisson equation for weak static fields, but also to guarantee that the equations be compatible with the law of energy-momentum conservation**.

On page 19R Einstein checked energy and momentum conservation for the resulting gravitational field equations in the case of weak fields. But he then discovered a problem. He began by writing:

"For the first approximation our additional condition is.

$$\sum_{\kappa} \gamma_{\kappa\kappa}\left(2\frac{\partial g_{i\kappa}}{\partial x_{\kappa}} - \frac{\partial g_{\kappa\kappa}}{\partial x_i}\right) = 0"$$

This is **the harmonic coordinate condition**.

Einstein then conjectured that the harmonic coordinate condition "can <u>perhaps</u> be decomposed into" two extra conditions.

The first condition is called by scholars **the "Hertz condition"**:

$$\sum_{\kappa} \gamma_{\kappa\kappa}\frac{\partial g_{i\kappa}}{\partial x_{\kappa}} = 0$$

because it was later mentioned by Einstein in a letter to Paul Hertz from August 22, 1915.[77]

This condition is related to the requirement of compatibility of the field equations and energy-momentum conservation,

And the second condition is related with compatibility of the field equations and the Newtonian limit, a condition on the trace of the weak field metric,

$$\sum \gamma_{\kappa\kappa}g_{\kappa\kappa} = \text{konst.}$$

**Entanglement of the Newtonian Limit and Energy-Momentum Conservation**

Einstein wanted to satisfy the energy-momentum conservation **and at the same time** to satisfy the Newtonian limit. On page 19L he introduced the harmonic coordinate condition to satisfy the weak field approximation. However, the Hertz condition was a troublesome condition for the above condition on the trace of the weak field metric: The trace of the metric tensor being constant was incompatible with the metric tensor diagonal $(-1, -1, -1, c^2)$ of a weak static gravitational field. In addition, it was incompatible with the field equations with the stress-energy tensor of matter for dust.

On page 19R the combination of the Hertz condition with the harmonic coordinate condition led to the unacceptable equation: $\sum \gamma_{\kappa\kappa} g_{\kappa\kappa} =$ konst. Energy conservation was found to hold, but the two conditions into which the harmonic coordinate condition had been split were not compatible with energy-momentum conservation; this meant an entanglement between the Newtonian limit and energy-momentum conservation.[78]

Einstein thus wrote at the bottom of page 19R, "<u>Beide</u> obige Bedingungen sind aufrecht zu erhalten" (Both the above conditions must be maintained).[79]

On page 20L Einstein again wrote the two conditions:

$$\sum \frac{\partial g_{i\kappa}}{\partial x_\kappa} = 0,$$

The Hertz condition, and the trace of the metric tensor:

$$\sum g_{kk} = 0,$$

and erased them by crossed lines, because the combination of these two conditions caused problems.[80]

Einstein used the harmonic and Hertz conditions to eliminate various terms from equations of broad covariance and looked upon the truncated equations of severely restricted covariance rather than upon the equations of broad covariance he started from as candidates for the fundamental field equations of his theory. Since coordinate conditions used in this manner are ubiquitous in the Zurich Notebook the scholars introduced a special name for them and called them *coordinate restrictions*.[81]

**Removing the "Hertz Coordinate Condition"**

Einstein then imposed the harmonic coordinate condition to reduce the Ricci tensor to the d'Alembertian acting on the metric in the weak-field case. The Hertz condition was added to make sure that the divergence of the stress-energy tensor vanishes in the weak-field case. But the combining of these two conditions implies that the trace of the metric has to vanish. Therefore, from the weak-field equations that Einstein obtained on page 19L (and written again on page 19R):

$$\sum \gamma_{\kappa\kappa} \frac{\partial^2 g_{im}}{\partial x^2_\kappa} = -\kappa \rho_0 \frac{dx_i}{ds} \frac{dx_m}{ds} g_{ii} g_{mm}.$$

it follows that the stress-energy tensor would be traceless. In order to avoid this problem Einstein modified these field equations and made their right hand side traceless. With his new weak-field equations Einstein managed to keep the stress-energy tensor and conservation principle. But this solved only part of the problem caused by the combination of the harmonic coordinate and the Hertz conditions. It takes care of the problem that a traceless metric would imply a traceless energy-momentum tensor, but it does not address a second problem, that a metric field of the form Einstein used to represent static fields was not traceless. Einstein thus crossed out his new field equations and modified again the weak-field equations. The harmonic coordinate condition was imposed and the Hertz condition was removed.[82]

Einstein was finally able to extract from the Ricci tensor linearized "gravitational equations" and recover the Newtonian limit successfully:

$$\sum g_{kk} = U,$$

$$\Delta\left(g_{11} - \tfrac{1}{2}U\right) = T_{11}, \quad \Delta\, g_{12} = T_{12}, \quad \dots \quad \Delta\, g_{14} = T_{14}.$$

Einstein's above modified weak-field equations had removed the need for the Hertz condition.

These weak field equations have exactly the same form as the weak-field equations found in Einstein's final general theory of relativity of November 25, 1915. The left-hand side is the linearized version of the Einstein tensor: $R_{\mu\nu} - (1/2)g_{\mu\nu}R$. There is no indication in the notebook that Einstein tried to find the exact equations corresponding to these weak-field equations.[83]

**Getting Rid of the Harmonic Coordinate Condition**

Again the contradiction between the coordinate conditions, which led to the troublesome additional condition for the linearized metric from page 19L; the latter was not satisfied by a metric of the form diagonal $(-1, -1, -1, c^2)$. On page 20L Einstein tried to avoid this problem by adding a trace term to the weak field equations, but then he confronted another problem: the metric of the form diagonal $(-1, -1, -1, c^2)$ was not anymore a solution of these **modified weak field equations**.

On page 21R Einstein returned to his static gravitational field equation from page 39L, and wrote "Statischer Spezialfall" ("static special case"). He considered the special case of the metric of static field. He expected to recover from his new metric tensor his static gravitational field equation: $c\Delta c - 1/2(\text{grad}c)^2 = \kappa c^2\sigma$ for the metric of the form $(-1, -1, -1, c^2)$. He wrote, "so müssen im statischen Felde $g_{11}$, $g_{22}$ etc. verschwinden" ("$g_{11}$, $g_{22}$ etc must vanish in the static field").[84] However, the flat metric of the form $(-1, -1, -1, c^2)$ represents the limit of special relativity, and the weak field equations no longer allow a solution with a metric of this form.[85]

But before giving up his new field equations Einstein wanted to check again whether his weak field metric is compatible with Galileo's experimental law of free fall and the equivalence principle. Einstein arrived at the conclusion that the metric of the form diagonal $(-1, -1, -1, c^2)$ is essential to Galileo's law.

Einstein wrote, "if the force is supposed to vary like the energy, then $g_{11}$, $g_{22}$ must vanish for the static field".[86] According to Galileo's experimental law of free fall, all bodies fall with the same acceleration in a given gravitational field. The **gravitational force** is then proportional to the **inertial mass**. According to special relativity, the **inertial mass** is proportional to **energy**. And therefore the field is represented by a flat metric of the form diagonal $(-1, -1, -1, c^2)$.[87]

Finally Einstein deleted the calculation of the "Static special case" and he wrote at the bottom of the page "Spezialfall wahrscheinlich unrichtig" ("special case probably incorrect)."[88] With that Einstein gave up the modified field equations from pages 19L-20L. At this early stage until 1915, he no longer considered field equations from the Ricci tensor with the help of the harmonic coordinate condition; nor did he examine modified weak-field equations, what was similar to the Einstein equations of November 1915 in linearized form: $\Sigma g_{kk} = U$.[89]

**The November Tensor**

Einstein was not ready to give up his attempt to extract the left hand-side of the field equations from the Riemann tensor. He took another approach to the problem of constructing a candidate generally covariant tensor from the Riemann curvature tensor. On page 22R Einstein wrote the heading "Grossmann". Thus perhaps at the suggestion of Grossmann, Einstein wrote another form of the Ricci tensor. This time, **the Ricci tensor was in terms of the Christoffel symbols** and their derivatives, rather than in terms of **the metric tensor** and its derivatives – as it appeared until then in the notebook. This was a fully covariant Ricci tensor in a form resulting from contraction of the Riemann tensor:

$$T_{il} = \sum_{\kappa l} \frac{\partial}{\partial x_l} \begin{Bmatrix} i\kappa \\ \kappa \end{Bmatrix} - \frac{\partial}{\partial x_\kappa} \begin{Bmatrix} il \\ \kappa \end{Bmatrix} + \begin{Bmatrix} i\kappa \\ \lambda \end{Bmatrix} \begin{Bmatrix} \lambda l \\ \kappa \end{Bmatrix} - \begin{Bmatrix} il \\ \lambda \end{Bmatrix} \begin{Bmatrix} \lambda \kappa \\ \kappa \end{Bmatrix},$$

where, $T_i = \frac{\partial lg\sqrt{G}}{\partial x_i}$ ,

Einstein wrote, "G ein Skalar ist", if G is a scalar, then the first term in the expansion of the Ricci tensor is itself a tensor of the first rank.

He divided the Ricci tensor into two parts – "Tensor 2. Ranges" (tensor of second rank) and "Vermutlicher Gravitations-Tensor" (presumed gravitational tensor):

$$T_{il} = \left(\frac{\partial T_i}{\partial x_l} - \sum \left\{\begin{matrix} il \\ \lambda \end{matrix}\right\} T_\lambda\right) - \sum_{\kappa l} \frac{\partial}{\partial x_\kappa} \left\{\begin{matrix} il \\ \kappa \end{matrix}\right\} - \left\{\begin{matrix} i\kappa \\ \lambda \end{matrix}\right\} \left\{\begin{matrix} l\lambda \\ \kappa \end{matrix}\right\}$$

The second term in the above equation which Einstein called "presumed gravitation tensor" is called by scholars the "November tensor":

$$\sum_{\kappa l} \frac{\partial}{\partial x_\kappa} \left\{\begin{matrix} il \\ \kappa \end{matrix}\right\} - \left\{\begin{matrix} i\kappa \\ \lambda \end{matrix}\right\} \left\{\begin{matrix} l\lambda \\ \kappa \end{matrix}\right\}.$$

Setting the November tensor equal to the energy-momentum tensor, multiplied by the gravitational constant κ, one arrives at the field equations of Einstein's first paper of November 4, 1915.[90]

**Comeback of the "Hertz Condition"**

On page 22L Einstein investigated the behavior of the November tensor. Einstein did not need any more the harmonic coordinate condition and he could impose the Hertz coordinate condition to eliminate all unwanted second-order derivative terms.[91] He wrote, "Weitere Umformung des Gravitationstensors" (further rewriting of the tensor of gravitation). The Hertz condition also ensured that the divergence of the linearized stress-energy tensor vanished. And so energy-momentum conservation law was satisfied, "Genugt, wenn $\sum \frac{\partial \gamma_{\kappa l}}{\partial x_l}$ verschwindet" (suffices, if it vanishes).[92]

**"Hertz Condition" Not Necessary**

At the bottom half of page 22R Einstein had arrived at a candidate for the left-hand side of the field equations extracted from the November tensor by imposing the Hertz condition. [93] But the conservation of energy-momentum could not be satisfied; and so on the next page 23L he went on to another novel method to eliminate terms with unwanted second-order derivatives of the metric and by which he could extract the Newtonian limit from the Riemann tensor. Einstein thus abandoned the Hertz condition:

"$\sum \frac{\partial \gamma_{\kappa\alpha}}{\partial x_\kappa} = 0$  sei = 0 ist nicht nötig." (not necessary). [94]

Einstein already accumulated coordinate conditions – or coordinate restrictions – to eliminate the terms from his equations. On pages 19L-23L Einstein extracted expressions of broad covariance from the Ricci tensor. He then truncated them by imposing additional conditions on the metric to obtain candidates for the left-hand side of the field equations that reduce to the Newtonian limit in the case of weak static fields.[95] But this model entangled the Newtonian limit and conservation of momentum-energy. One page after the other in the Zurich Notebook, he turned from one candidate equation to another to find the suitable left hand side of the field equations that would be compatible with energy-momentum conservation. The equations satisfied conservation of energy-momentum in the weak field level, but the source term – stress-energy of matter – of the gravitational field was incompatible with what he had obtained for the Newtonian limit.[96]

**Field Equations Found Through Conservation of Energy and Momentum**

On page 24R Einstein tried to extract yet another candidate for the left-hand side of the field equations. **He did not extract the candidate from the Ricci tensor** while imposing coordinate conditions. He established field equations while starting from the requirement of the conservation of momentum and energy. The equations could be covariant with respect to linear transformations, and they satisfied both the Newtonian limit and conservation of momentum-energy.

Einstein checked this candidate using the rotation metric. He thought that his expression vanished for the rotation metric, a necessary condition for the rotation metric to be a solution of the vacuum field equations. This was a mistake, but Einstein was to discover this much later (October 1915). His expression vanishing for the rotation metric could signify to Einstein that the new field equations satisfied the relativity principle and the equivalence principle. Einstein came to believe that his expression vanishes for the rotation metric because of a sign error in $\gamma_{12}$ and $\gamma_{21}$.[97]

He would do quite the same sign error and come to a similar belief with respect to the rotation metric a year later in the "Einstein-Besso manuscript".[98] He there checked whether the rotation metric is a solution of the newly "Entwurf" equations he was developing.[99]

Rather than now making the correction of the error above made, Einstein on pages 25L and 25R was trying to find a way to recover the new field equations from the November tensor of page 22R which he wrote again on top of page 25L.[100]

**Connecting Between New Field Equations and the November Tensor**

Einstein still hoped to connect his new field equations found through energy-momentum considerations to the November tensor of page 22R. On pages 25L-25R, he explicitly tried to recover field equations along this argument from this tensor. At some point he came to reject his efforts at recovering his new field equations found through energy-momentum considerations from the November tensor, and he abandoned general covariance. He failed to connect the November tensor to his new field equations of page 24R. Einstein then wrote in the lower left corner of page 25R "Unmöglich" (impossible).[101]

**The "Entwurf" Field Equations**

This brought Einstein on the very next pages, 26L and 26R, straight to the field equations, the "Entwurf" equations, which he also established by the same method: through energy-momentum considerations. The problem remained whether these equations were covariant enough to enable extending the principle of relativity for accelerated motion and to satisfy the equivalence principle.

Under the title "System der Gleichungen für Materie" (System of Equations for Mater), and "Ableitung der Gravitationsgleichungen" (Derivation of the gravitational equations) Einstein derived gravitational field equations of limited covariance that **were not derived from the Riemann tensor**.[102] These equations spread over two facing pages, 26L and 26R and are displayed with a neatness and order rare among the other pages of the notebook, suggesting that they were transcribed from another place (probably from Einstein's and Grossmann's joint 1913 "Entwurf" paper) after the result was known. Einstein ended his gravitation calculations at page 26R with the left-hand side of the "Entwurf" gravitation tensor.[103]

As seen from examining the Zurich Notebook, three years before November 1915, Einstein had written on page 22R the November tensor, when he considered the Ricci tensor as a possible candidate for the left hand-hand side of his field equations. Einstein got so close to his November 1915 breakthrough at the end of 1912, that he even considered on page 20L another candidate – albeit in a linearized form – which resembles the final version of the November 25, 1915 field equation of general relativity $[R_{\mu\nu} - (1/2)g_{\mu\nu}R]$. Einstein therefore first wrote down a mathematical expression close to the correct field equation and then discarded it, only to return to it more than three years later.

Why did Einstein reject in 1912-1913 gravitational field equations of much broader covariance, only to come back to these field equations in November 1915? Einstein believed that the special principle of relativity for uniform motion could be generalized to arbitrary motion if the field equations possessed the mathematical property of general covariance (that is, a form which remained unchanged under all coordinate transformations). If the principle of relativity is generalized then the equivalence principle is satisfied. Accordingly, Einstein examined candidates for generally covariant field equations.

Einstein's earlier work on static fields led him to conclude (on page 21R) that in the weak field approximation, the spatial metric of a static gravitational field must be flat.[104] This statement appears to have led him to reject the Ricci tensor of page 22R, and fall into the trap of "Entwurf" limited generally covariant field equations. The "Entwurf" gravitational equations were thus incompatible with a general principle of relativity. Einstein said in the introduction to his November 4, 1915 paper, [105]

"For these reasons, I completely lost trust in my established field equations, and looked for a way to limit the possibilities in a natural manner. Thus I arrived back at the demand of a broader general covariance for the field equations, from which I parted, though with a heavy heart, three years ago when I worked together with my friend Grossmann. As a matter of fact, we then have already come quite close to the solution of the problem given in the following".

Since Einstein thought that the Ricci tensor should reduce in the limit to his static gravitational field theory from 1912 and then to the Newtonian limit, if the static spatial metric is flat, then this prevented the Ricci tensor from representing the gravitational potential of any distribution of matter, static or otherwise.[106]

John Stachel explained that Einstein attempted to formulate in the best way he could his physical insights about gravitation and relativity already, and incorporate them in the equivalence principle. Einstein's attempt was hampered by the absence of the appropriate mathematical concepts. Until 1912 Einstein lacked the Riemanian geometry and the tensor calculus as developed by the turn of the century, i.e., based on the concept of the metric tensor; and after 1912 when he was using these, he then lacked more advanced mathematical tools (the affine connection)[107]; these could be later responsible for inhibiting him for another few years.[108]

Stachel concludes that in the absence of the affine approach, more-or-less heuristic detours through the weak field, fast motion (i.e., special-relativistic) limit followed by a slow motion approximation basically out of step with the fast-motion approach, had to be used to "obtain" the desired Newtonian results.[109]

Judged from the historical point of view of his time, Einstein did not make a mistake, because he lacked the appropriate mathematical tools to correctly taking the Newtonian limit of general relativity. Actually with hindsight the story is more complicated. What was eventually mere coincidence for Einstein would later turn to be a consequence derived by new mathematical tools, the affine connection, which was invented after Einstein had arrived at generally covariant field equations.[110]

Indeed after 1916 the mathematical tools were elaborated. Stachel summarizes,[111]

"Had he known about the connection representation of the inertio-gravitational field, he would have been able to see that the spatial metric can go to a flat Newtonian limit, while the Newtonian connection remains non-flat without violating the compatibility conditions between metric and connection. As it was, […], he was amazed to find that the spatial metric is non-flat".

## 2 The "Entwurf" Theory – Zurich 1913

In Einstein and Grossmann's "Entwurf" paper Grossmann wrote the mathematical part and Einstein wrote the physical part. The paper was first published in 1913 by B. G. Teubner (Leipzig and Berlin). And then it was reprinted with added "Bemarkungen" (remark) in the *Zeitschrift für Mathematik und Physik* in 1914. The "Bemarkungen" was written by Einstein and contained the well-known "Hole Argument".[112]

Einstein and Grossmann developed a new theory of gravitation which was based on absolute differential calculus. They first established the system of equations for material processes when the gravitational field was considered as given. These equations were covariant with respect to arbitrary substitutions of the space-time coordinates. After establishing these equations, they went on to establish a system of equations which were regarded as a generalization of the Poisson equation of Newton's theory of gravitation. These equations determine the gravitational field, provided that the material processes are given. In contrast to the equations for material processes, Einstein and Grossmann could not demonstrate general covariance for the latter gravitational equations. Namely, their derivation was assumed – in addition to the conservation laws – only upon the covariance with respect to *linear* substitutions, and not upon arbitrary transformations.

Einstein felt that this issue was crucial, because of the equivalence principle. His theory depended upon this principle: all physical processes in a gravitational field occur just in the same way as they would without the gravitational field, if one related them to an appropriately accelerated (three-dimensional) coordinate-system. This principle was founded upon a fact of experience, that of the equality of inertial and gravitational masses.

Einstein's desire was that acceleration-transformations – nonlinear transformations – would become permissible transformations in his theory. In this way transformations to accelerated frames of reference would be allowed and the theory could generalize the principle of relativity for uniform motions. Einstein thus understood that it was desirable to look for gravitational equations that are covariant with respect to arbitrary transformations.

## 2.1 The Equivalence Principle

The "Entwurf" paper is divided into two parts: a physical part written by Einstein and a mathematical part written by Grossmann. Einstein begins the physical part with his new law of nature: the equality of inertial and gravitational masses. From this Einstein was led to the hypothesis that, from a physical point of view an (infinitesimally extended, homogeneous) gravitational field can be completely replaced by a state of acceleration of the reference system. The law of nature is the equality of inertial and gravitational masses and the equivalence principle is a hypothesis.[113]

Einstein explained the "equivalence hypothesis" by using the predecessor of the elevator thought experiment: an observer enclosed in a box can in no way decide whether the box is at rest in a static gravitational field, or whether it is in accelerated motion, maintained by forces acting on the box, in space that is free of gravitational fields.[114]

## 2.2 The Static Gravitational Field

In section §1 Einstein presented the equations of motion of the material point in the static gravitational field. He did not abandon his 1912 static fields theory (section §1 of the "Entwurf" paper), but rather developed it, and it became a starting point and a limiting case of the new theory, and also a kind of inevitable trap of his theory.

Einstein stated that he had shown in previous papers that the equivalence hypothesis leads to the consequence that in a static gravitation field the velocity of light c depends on the gravitational potential. This led him to think that the usual (special) theory of relativity provides only an approximation to reality. It should apply only in the limiting case where differences in the gravitational potential in the space-time region under consideration are not too great. He found that c should not be conceived as a constant. It rather should be a function of the special coordinates that represent a measure for the gravitational potential. Einstein thus arrived at equations of motion for the material point and to the conclusion that in the static case c plays the role of the gravitational potential.[115]

## 2.3 Einstein's Static Field and Mach's Idea

Einstein ended section §1 with the conclusion that the momentum and kinetic energy are inversely proportional to c. Or, the inertial mass is m/c and independent of the gravitational potential.[116] This conforms to Mach's idea that inertia has its origin in an interaction between the mass point under consideration and all of the other mass points. Einstein explained that if other masses are accumulated in the vicinity of the mass point, the gravitational potential c decreases. And then the quantity m/c increases which is equal to the inertial mass. In the static fields theory Einstein presented the predecessor to Mach's principle.[117]

## 2.4 The Metric Degenerates in Static Fields to $g_{44} = c^2$ and Space is Flat

In section §2 Einstein was dealing with correspondence to his static gravitational fields theory. Einstein planned to generalize the principle of relativity in such a way that the theory of static gravitational fields from section §1 will be contained in his new theory presented in this paper as a special case.

Here Einstein was trying to do what he has done in the Zurich Notebook on page 21R.[118] And so he concluded that in the general case, the gravitational field is characterized by ten space-time functions (of the metric tensor); $g_{\mu\nu}$ are functions of the coordinates $(x_\nu)$.[119] In the case of the usual (special) theory of relativity this reduces to $g_{44} = c^2$, where c denotes a constant. And Einstein said that the same degeneration occurs in the static gravitational field of the kind he considered in section §1, except that in the latter case, this reduces to a single potential $g_{44} = c^2$, where $g_{44} = c^2$ is a function of spatial coordinates, $x_1, x_2, x_3$.[120]

## 2.5 Equations of Motion of a Mass Point in an Arbitrary Gravitational Field

Einstein then established the equations of motion of a material point in an arbitrary gravitational field. He followed his calculations from the Zurich Notebook at the top of page 05R. There he had started with the equations of motion of a point particle in a metric field from an action principle.[121]

In section §1 he dealt with the system K, in which the gravitational field was static. He presented another space-time system K' in which the gravitational field was arbitrary. Following Max Planck in 1906, with respect to K', Einstein derived the equations of motion of the freely moving material point from a variation principle in its Hamiltonian form:

$$\delta\{\textstyle\int ds\} = \delta\{\textstyle\int \sqrt{(-\,dx^2 - dy^2 - dz^2 + c^2 dt^2)}\} = 0$$

where,

$$ds^2 = \Sigma_{\mu\nu} g_{\mu\nu} dx_\mu dx_\nu$$

After further calculations of the momentum of the material point, Einstein concluded that the quantities $g_{\mu\nu}$ form a covariant tensor of the second rank, which he called the "covariant fundamental tensor". This tensor determines the gravitational field. He further arrived at the results that the momentum and energy of the material point form together a covariant tensor of the first rank, i.e. a covariant vector. Subsequently he referred the reader to Grossman's mathematical part for further explanations of this issue.[122]

## 2.6 The "Natural Interval"

Already in the Zurich Notebook on page 05R the equations of motion of the point particle were written in the form of Euler-Lagrange. The action integral was the proper length of the particle's worldline. What is a worldline from a physical point of view? Einstein was looking for physical meaning to the mathematical quantities and so in Section §3 he dealt with the significance of the fundamental tensor of the $g_{\mu\nu}$ for the measurement of space and time. He said that from what he laid down in section §2 it is obvious that **one can use measuring rods and clocks in much the same way as one can do in the usual (special) theory of relativity**. Einstein thus sought for the **physical meaning** – the measurability – of the space-time quantities $x_1$, $x_2$, $x_3$, $x_4$.

Sometime around November 18, 1915, Einstein found out that in general relativity one could not use measuring rods and clocks in the same way as one would do in special relativity, because "time and space are deprived of the last trace of objective reality".[123] Thus space and time coordinates have no meaning in general relativity.

In his 1913 "Entwurf" theory Einstein considered two infinitely close space-time points. ds **possesses a physical meaning** that is independent of the chosen reference system. Einstein assumed that ds is **the "naturally measured" interval** between the two space-time points, or the square of the four-dimensional interval between two infinitely close space-time points. It is measured by means of a rigid body that is not accelerated in a system which is introduced by means of linear transformations with respect to the immediate vicinity system of the point $(dx_1, dx_2, dx_3, dx_4)$, and by means of unit measuring rods and clocks at rest relative to it.

For given $dx_1, dx_2, dx_3, dx_4$, the natural interval that corresponds to these differentials can be determined only if one knows the quantities $g_{\mu\nu}$ that determine the gravitational field. Or the gravitational field influences the measuring bodies and clocks in a determinate manner.

Einstein then concluded that from the fundamental equation (the line element):

$$ds^2 = \Sigma_{\mu\nu} g_{\mu\nu} dx_\mu dx_\nu$$

one sees that in order to fix the physical dimensions of the quantities $g_{\mu\nu}$ and $x_\nu$, another stipulation is required. ds has the dimensions of length, and so does $x_\nu$ and $x_4$ ("time"), and he did not ascribe any physical dimension to the quantities $g_{\mu\nu}$.[124]

## 2.7 The Conservation Law of Energy Momentum for Matter

In section §4 Einstein started with continues incoherent masses moving in arbitrary gravitational fields; the cloud of pressureless dust particles in the presence of a gravitational field from page 05R of his Zurich Notebook. For this case Einstein applied the conservation of momentum law. He arrived at a form for the conservation law of energy momentum for matter, which was already found in on page 05R[125]:

$$\sum_{\mu\nu} \frac{\partial}{\partial x_\nu}(\sqrt{-g} \cdot g_{\sigma\mu}\Theta_{\mu\nu}) - \frac{1}{2}\sum_{\mu\nu} \sqrt{-g} \cdot \frac{\partial g_{\mu\nu}}{\partial x_\sigma}\Theta_{\mu\nu} = 0$$

The first three of these equations ($\sigma = 1, 2, 3$) express the momentum law, and the last one ($\sigma = 4$) the energy law.[126]

These equations are covariant with respect to arbitrary substitutions. And Einstein again referred to Grossmann's mathematical part.[127]

## 2.8 The Contravariant Stress-Energy Tensor

Einstein called the tensor $\Theta_{\mu\nu}$ the "(contravariant) *stress-energy tensor of the material flow*". He then said that he ascribed to the above equation a validity that goes far beyond the special case of the flow of incoherent masses. The above equation represents in general the energy balance between the gravitational field and an arbitrary material process; one had only to substitute for $\Theta_{\mu\nu}$ the stress-energy tensor corresponding to the material system under consideration. Einstein further explained that the first term in the equation contains the space derivatives of the stresses or of the density of the energy flow, and the time derivatives of the momentum density or the energy density. The second term is an expression for the effects exerted by the gravitational field on matter.[128]

## 2.9 The Field Equations

Einstein now advanced in section §5 to the differential equations of the gravitational field itself. Einstein started from his new tensor, $\Theta_{\mu\nu}$, the stress-energy tensor, for the material processes. He then asked: What differential equations permit us to determine the quantities $g_{ik}$, that is, the gravitational field? He wanted to find equations from which he would be able to calculate the quantities $g_{ik}$ when the quantities of the material processes are known.

Einstein sought the generalization of Poisson's equation:

$\Delta\varphi = 4\pi k\rho.$

The generalization Einstein was seeking would likely have the form:

$\chi \cdot \Theta_{\mu\nu} = \Gamma_{\mu\nu},$

where $\chi$ is a constant, analogous to the Newtonian gravitation constant G, $\Theta_{\mu\nu}$ is analogous to the source mass-density of the Poisson equation $\rho$, and $\Gamma_{\mu\nu}$ is a second-rank contravariant tensor derived from the fundamental metric tensor $g_{\mu\nu}$ by differential operations.[129]

In line with the Newton-Poisson equation Einstein now presumed the above equations would **be second order**. But then he confronted the problem of being unable to find a differential expression $\Gamma_{\mu\nu}$ that is a generalization of $\Delta\varphi$ and that proves to be a tensor with respect to arbitrary transformations. And again Einstein referred to Grossmann's mathematical part for this defect.[130]

As quoted in extension further below, Grossman explained that the covariant differential tensor of second rank – the Ricci tensor – would have been the natural candidate for $\Gamma_{\mu\nu}$. But it turns out that in the special case of the infinitely weak, static gravitational field this tensor does *not* reduce to the expression $\Delta\varphi$.

Einstein said that it cannot be negated that the final, exact equations of gravitation could be of higher than second order. And then there will be a possibility that the perfectly exact differential equations of gravitation would be after all covariant with respect to arbitrary transformations. But given the present state of his knowledge of the physical properties of the gravitational field, the attempt to discuss such possibilities would be premature. "Therefore, given the limitation of the second order, we must forgo establishing gravitational equations that are covariant with respect to arbitrary transformations. It should be emphasized, incidentally, that we have no evidence for general covariance of the gravitational equations".[131]

Einstein decided to follow the spirit of Poisson's equation, and to give up searching for generally covariant field equations, equations of gravitation that are covariant with respect to arbitrary transformations.

Following the correspondence principle, the Newtonian limit, the field equations are covariant only with respect to a particular group of transformations, which group was as yet unknown to Einstein at this stage. But given the usual (special) theory of relativity, Einstein reasoned that it was naturally to assume that the transformation group he was seeking also includes the linear transformations. Hence he required that $\Gamma_{\mu\nu}$ be a tensor with respect to any or arbitrary linear transformations. Einstein could now obtain an expression for a covariant tensor of the second rank with respect to the linear transformations.[132]

## 2.10 The Contravariant Form of the Field Equations

Einstein wrote the conservation law for the gravitational field: [133]

$$\sum_{\mu\nu} \frac{\partial}{\partial x_\nu}(\sqrt{-g} \cdot g_{\sigma\mu} \cdot \vartheta_{\mu\nu}) - \frac{1}{2}\sum_{\mu\nu}\sqrt{-g} \cdot \frac{\partial g_{\mu\nu}}{\partial x_\mu} \cdot \vartheta_{\mu\nu} = -\frac{1}{2\chi} \cdot \sum_{\mu\nu}\sqrt{-g} \cdot \frac{\partial g_{\mu\nu}}{\partial x_\sigma} \cdot \Delta_{\mu\nu}(\gamma).$$

$\Delta(\gamma)$ is given by:

$$\Delta_{\mu\nu}(\gamma) = \sum_{\alpha\beta}\frac{1}{\sqrt{-g}} \cdot \frac{\partial}{\partial x_\alpha}\left(\gamma_{\alpha\beta}\sqrt{-g} \cdot \frac{\partial\gamma_{\mu\nu}}{\partial x_\beta}\right) - \sum_{\alpha\beta\tau\rho}\gamma_{\alpha\beta}g_{\tau\rho}\frac{\partial\gamma_{\mu\tau}}{\partial x_\alpha}\frac{\partial\gamma_{\nu\rho}}{\partial x_\beta}.$$

And, $\vartheta_{\mu\nu}$ is the contravariant stress-energy tensor of the gravitational field, which enters the conservation law for the gravitational field in exactly the same way as the tensor $\Theta_{\mu\nu}$ of the material process enters the conservation law for this process. It is given by:

$$-2\chi \cdot \vartheta_{\mu\nu} = \sum_{\alpha\beta\tau\rho}(\gamma_{\alpha\mu}\gamma_{\beta\nu}\frac{\partial g_{\tau\rho}}{\partial x_\alpha} \cdot \frac{\partial\gamma_{\tau\rho}}{\partial x_\beta} - \frac{1}{2}\gamma_{\mu\nu}\gamma_{\alpha\beta}\frac{\partial g_{\tau\rho}}{\partial x_\alpha}\frac{\partial\gamma_{\tau\rho}}{\partial x_\beta}).[134]$$

See page 28L of the Zurich Notebook. [135]

Einstein arrived at the (contravariant) form of the gravitational equations:

$\chi \cdot \Theta_{\mu\nu} = \Gamma_{\mu\nu}$,

$\Delta_{\mu\nu}(\gamma) = \chi(\Theta_{\mu\nu} + \vartheta_{\mu\nu})$.

The field equations show that the stress-energy tensor $\vartheta_{\mu\nu}$ of the gravitational field acts as a field generator in the same way as the tensor $\Theta_{\mu\nu}$ of the material process. [136]

Einstein next showed that the conservation laws hold for the matter and the gravitational field together by including them in the equation: [137]

$$\sum_{\mu\nu} \frac{\partial}{\partial x_\nu} \{\sqrt{-g} \cdot g_{\sigma\mu}(\Theta_{\mu\nu} + \theta_{\mu\nu})\} = 0$$

## 2.11 The Covariant Form of the Field Equations

In their covariant form the gravitational equations are: [138]

$-D_{\mu\nu}(g) = \chi(t_{\mu\nu} + T_{\mu\nu})$,

where, $t_{\mu\nu}$ is the covariant stress-energy tensor of the gravitational field:[139]

$$-2\chi \cdot t_{\mu\nu} = \sum_{\alpha\beta\tau\rho} \left( \frac{\partial g_{\tau\rho}}{\partial x_\mu} \frac{\partial \gamma_{\tau\rho}}{\partial x_\nu} - \frac{1}{2} g_{\mu\nu} \gamma_{\alpha\beta} \frac{\partial g_{\tau\rho}}{\partial x_\alpha} \frac{\partial \gamma_{\tau\rho}}{\partial x_\beta} \right).$$

and $T_{\mu\nu}$ is the covariant stress-energy tensor of matter:

$$T_{\mu\nu} = \sum_{\alpha\beta} g_{\mu\alpha} \, g_{\nu\beta} \, \Theta_{\alpha\beta}.$$

D(g) is given by:

$$D_{\mu\nu}(g) = \sum_{\alpha\beta} \frac{1}{\sqrt{-g}} \cdot \frac{\partial}{\partial x_\alpha} \left( \gamma_{\alpha\beta} \sqrt{-g} \, \frac{\partial g_{\mu\nu}}{\partial x_\beta} \right) - \sum_{\alpha\beta\tau\rho} \gamma_{\alpha\beta} \gamma_{\tau\rho} \frac{\partial g_{\mu\tau}}{\partial x_\alpha} \frac{\partial g_{\nu\rho}}{\partial x_\beta}.$$

Einstein ended section §5 by writing the conservation law of energy momentum for matter in covariant form:[140]

$$\sum_\nu \frac{\partial}{\partial x_\nu} \{\sqrt{-g} \cdot \gamma_{\sigma\mu}(T_{\mu\nu} + t_{\mu\nu})\} = 0.$$

## 2.12 Grossmann's Mathematical Part

Grossmann started his mathematical part with the invariance of the line element:

$ds^2 = \Sigma_{\mu\nu} g_{\mu\nu} dx_\mu dx_\nu$

Grossmann said that the mathematical tools concerning this line element are found in Christoffel's paper from 1869, "On the Transformation of the Homogeneous differential Forms of the Second Degree".[141]

In sections §1 to §3 Grossmann gave a preliminary introduction to tensor analysis.[142] In section §4 he started with the mathematical supplements to Einstein's physical part section §4. He first supplied a proof of the covariance of the momentum-energy equations, which Einstein presented in the physical part. He arrived at a covariant equation and concluded, "*The divergence of the (contravariant) stress-energy tensor of the material flow, or of the physical process vanishe*s".[143]

Grossmann next complemented mathematically Einstein's section §5. The problem of constructing the differential equations of the gravitational field is connected with the differential tensors that are given with a gravitational field. Grossmann said that the complete system of these differential tensors (with respect to arbitrary transformations) goes back to the covariant differential tensor of fourth rank found in Riemann's and Christoffel's works. He wrote Riemann's differential tensor.[144]

Subsequently, Grossmann wrote the complementary mathematical argument to Einstein's physical argument in section §5, [145]

"The extraordinary importance of these conceptions for the power of *differential geometry* of a line element that is given by its manifold makes it a priori probable that these general differential tensors may also be of importance for the problem of the differential equations of a gravitational field. It is possible, in fact, at first to specify a covariant differential tensor of second rank and second order for $G_{im}$ to specify which one could enter into these equations [the Ricci tensor…]

But it turns out that in the special case of the infinitely weak, static gravitational field this tensor does *not* reduce to the expression $\Delta\varphi$. We must therefore leave open the question to what extent the general theory of the differential tensors associated with a gravitational field is connected with the problem of the gravitational equations. Such a connection would have to exist, provided that the gravitational equations are to permit arbitrary substations; but in that case, it seems that it would be impossible to find second-order differential equations. On the other hand, if it were determined that the gravitational equations permit only a certain group of transformations, then it would be understandable if one could not manage to provide the differential tensors by the general theory. As has been explained in the physical part, we are not able to take a stand on these questions".

Grossmann appeared to have been influenced by Einstein's conception that, in the weak field approximation, the spatial metric of a static gravitational field must be flat. This prevented the Ricci tensor from representing the gravitational potential.[146]

## 3 The Hole Argument – 1913-1914

### 3.1 Dissatisfaction with Limited Covariant Field Equations

**On August 14, 1913**, Einstein was dissatisfied with his "Entwurf" field equations, for he wrote to Lorentz, [147]

"But, unfortunately, this matter hooks me so much, that my confidence in the admissibility of the theory is still shaky. The Entwurf is satisfactory so far, insofar as it concerns the effect of the gravitational field on other physical processes. For the absolute differential calculus permits the setting up of equations here that are covariant with respect to arbitrary substitutions. The gravitational field ($g_{\mu\nu}$) seems to be the skeleton, so to speak, on which everything hangs. *But unfortunately, the* *gravitation equations themselves do not possess the property of general covariance*. Only their covariance with respect to *linear* transformations is certain. Now, however, all of the confidence in the theory rests on the conviction that an acceleration of the reference system is equivalent to a gravitational field. So if not all of the system of equations of the theory, and thus also the gravitational equations permit other than linear transformations, then the theory refutes its own starting point; then it stands in the air".

Two days later Einstein wrote Lorentz and found that assuming the law of momentum and energy conservation, his gravitational equations are never absolutely covariant, [148]

"I also found out yesterday to my greatest satisfaction the opposite to the doubts regarding the gravitation theory, which appear in my last letter, as well as those expressed in the paper. The matter seems to me solved as follows. The expression for the energy principle for matter & gravitation field taken together […] starting out from this assumption, I set up equations [… the gravitational equations]. Now, however, a consideration of the general differential operators of the absolute differential calculus shows that such an equation is never constructed absolutely covariant. As we thus postulated the existence of such an equation, we tacitly specialized the choice of the reference system. We restricted ourselves to the use of such reference systems with respect to which the law of momentum and energy conservation holds in this form. It appears that if one privileges such reference systems, then only more general linear transformations remain as the only right choice.

Hence, in a word: *By postulating the conservation law, one arrives at highly particular choice of the reference system and the admissible substitutions*.

Only now, does the theory give me pleasure after this ugly dark spot seems to have been eliminated".

**On September 9, 1913**, in Einstein's lecture before the annual meeting of the Naturforschende Gesellschaft in Frauenfeld, "Physikalische Grundlagen einer Gravitationstheorie" ("Physical Foundations of a Theory of Gravitation"), Einstein spoke of "a general consideration", [149]

"It has been possible to demonstrate by a general consideration that equations that can be covariant determine the gravitational field with respect to non arbitrary substitutions. This fundamental knowledge is therefore especially remarkable because all other physical equations, […] are not covariant with respect to arbitrary, but only with respect to linear transformations. We will therefore also have to request for the desired field equations only the covariance with respect to linear transformations. Added to these considerations, it has been found that we can completely perform certain equations that must emerge from the equations in the special case as an approximation of the Poisson equation". [150]

The general consideration could be the explanation presented to Lorentz on August 16 or a new consideration. Einstein ended his paper by saying, [151]

"By the theory outlined an epistemological defect is removed, which is inherent not only to the original theory of relativity, but also to the Galilean mechanics, and had been emphasized especially by E. Mach. […] It has been shown that in fact the [gravitational] equations indicate the behavior of the inertial resistance, which we can denote as the inertia of relativity. This fact is one of the main pillars of the outlined theory".

Although Einstein felt that he managed for the time being to remove the epistemological defect, he could not deal with the mathematical defect of general covariance.

Immediately after Einstein's talk, Grossman presented his talk in the same Naturforschende Gesellschaft in Frauenfeld, "Mathematische Begriffsbildungen zur Gravitationstheorie" ("Mathematical Concepts of Gravitational Theory"). Grossman opened his talk by mentioning the works of Christoffel from 1869, Levi-Civita and Ricci from 1901, and the developments since then to the theory of invariants and the new treatments by Minkowski, Sommerfeld and Laue. [152] However, Grossman did not solve the mathematical defect in Einstein's theory: the general covariance problem of the gravitational field equations.

Two weeks later, **on September 23, 1913**, Einstein attended the $85^{th}$ congress of the German natural Scientists and Physicists in Vienna. There he presented another talk, "Zum gegenwärtigen Stande des Gravitationsproblems" ("On the Present State of the Problem of Gravitation") pertaining to his "Entwurf" theory. He also engaged in a dispute after this talk with scientists who opposed to his theory. A text for this lecture with the discussion was published in the December volume of the *Physikalische Zeitschrift*. Einstein rewrote in section §7 of his Vienna paper the gravitational equations he had obtained in the "Entwurf" paper. [153] And thus there was little new under the sun in the Vienna talk. Before presenting these equations he wrote, [154]

"The whole problem of gravitation would therefore be solved satisfactorily, if one were also able to find such equations *covariant with respect to any arbitrary transformations* that are satisfied by the quantities $g_{\mu\nu}$ that determine the gravitational field itself. We have not succeeded in solving that problem in this manner".

But after the words "in this manner" Einstein added the following footnote, which indicated that he did arrive at some new idea, "2) In the last few days, I have found a proof that such a generally covariant solution cannot exist at all". [155]

This footnote appeared in the printed version of the Vienna lecture – published **on December 15, 1913**. Hence says Stachel the footnote could be added only later, after September 1913. If it was added just before December 1913, then a short time before December 1913, **in November 1913**, Einstein arrived at an ingenious idea. [156]

Einstein told Ludwig Hopf **on November 2, 1913**:

"I am now very happy with the gravitation theory. The fact that the gravitational equations are not generally covariant, which bothered me some time ago, has proved to be unavoidable; it can easily be proved that a theory with generally covariant equations cannot exist if it is required that the field be mathematically completely determined by matter".[157]

And the proof was **the Hole Argument**.

### 3.2 The Hole Argument

"Hole argument" is the English translation of the German phrase "*Lochbetrachtung*". Einstein developed this argument against the possibility of generally-covariant equations for the gravitational field. The argument was first published in the "Bemarkungen" (remarks"), which forms the addendum to the 1913 "Entwurf" paper.[158]

The Hole Argument reappeared again in Einstein's paper, "Prinzipielles zur verallgemeinerten Relativitätstheorie und Gravitationstheorie" ("On the Foundations of the Generalized Theory of Relativity and the Theory of Gravitation"), published in January 1914 in *Physikalische Zeitschrift*. Einstein was already defensive at this stage, because his failure to offer generally covariant field equations was a great worry and embarrassment for him.[159] Einstein wrote, " 'Very true', thinks the reader, 'but the fact the Messrs. Einstein and Grossmann are not able to give the equations for the gravitational field in generally covariant form is not a sufficient reason for me to agree to a specialization of the reference system'. But there are two weighty arguments that justify this step, one of them of logical, the other one of empirical provenance". The logical argument is the Hole Argument.[160]

These two versions are the first versions of the Hole Argument.

## 4 The Einstein-Grossman second "Entwurf" paper

Around March 1914 – Einstein was about to leave Zurich and start his Berlin period. His collaboration with Grossmann was going to end. Before Einstein left he wrote his last joint paper with Grossmann that included just another excuse for not presenting generally covariant field equations, "Kovarianzeigenschaften der Feldgleichungen der auf die verallgemeinerte Relativitätstheorie gegründeten Gravitationstheorie" ("Covariance Properties of the Field Equations of the Theory of Gravitation Based on the General Theory of Relativity").[161] The paper was published in May 1914 when Einstein was already in Berlin.

### 4.1 A New Condition that saves the 'Entwurf" theory

In section §2 of the paper Einstein brought a new "simple consideration" according to which $g_{\mu\nu}$ that characterize the gravitational field could not be completely determined by generally-covariant equations. Einstein's new argument not only justified the restricted covariance of the field equations; it even supplied reasoning for why generally covariant field equations would be unacceptable. [162]

Einstein first explained the argument to his close friend Michele Besso on March 10, 1914. Einstein wrote Besso that his novelty in the gravitation theory was the following. From the gravitation equations:

$$(1) \quad \sum_{\alpha\beta\mu} \frac{\partial}{\partial x_\alpha}\left(\sqrt{-g}\,\gamma_{\alpha\beta}g_{\sigma\mu}\frac{\partial\gamma_{\mu\nu}}{\partial x_\beta}\right) = \kappa(T_{\sigma\nu} + t_{\sigma\nu})$$

And from the conservation law it follows that:

$$(2) \quad \sum_{\alpha\beta\mu\nu} \frac{\partial}{\partial x_\nu}\frac{\partial}{\partial x_\alpha}\left(\sqrt{-g}\,\gamma_{\alpha\beta}g_{\sigma\mu}\frac{\partial\gamma_{\mu\nu}}{\partial x_\beta}\right) = 0.$$

Einstein wrote these in short: $B\sigma = 0$.

These are 4 third order equations for the $g_{\mu\nu}$ (or $\gamma_{\mu\nu}$) which can be conceived as the conditions for the special choice of the reference system. Einstein told Besso that by means of a simple calculation he can prove that the gravitation equations hold for every reference system that is adapted to this condition.

Einstein thus concluded, [163]

"This shows that there exists acceleration transformations of varied kinds, which transform the equations to themselves (e.g. also rotation), so that the equivalence hypothesis is preserved in its original form, even to an unexpectedly large extent".

In section §1 "An Stelle der Gravitationsgleichung (21) bzw. (18) des 'Entwurfes'" of Einstein and Grossmann's paper, Einstein wrote the above gravitational equations (1).[164] In section §2 he also obtained the second equation (2) appearing in Einstein's letter to Besso.[165]

Einstein was so happy that he wrote Besso, "Now I am perfectly satisfied and no longer doubt the correctness of the whole system, regardless of whether the observation of the solar eclipse will be successful or not. The logic of the thing is too evident".

Einstein said in his 1914 joint paper with Grossmann, "*the gravitational equations established by us are generally covariant just to the degree, which is possible under the condition that the fundamental tensor $g_{\mu\nu}$ is completely determined. It follows in particular that the gravitational equations are covariant with respect to acceleration transformations (i.e., nonlinear transformations) of about various kinds*".[166]

## 4.2 Falling Deeper into the Hole

In addition to the above new "simple consideration" brought in section §2 of the paper, Einstein wrote,[167]

"We want to show now at first that, completely independent of the gravitational equations that we established, a complete determination of the fundamental tensor $\gamma_{\mu\nu}$ of the gravitational field at a given $\Theta_{\mu\nu}$ by a generally-covariant system of equations is impossible.

Namely, we can prove that if a solution for the $\gamma_{\mu\nu}$ at a given $\Theta_{\mu\nu}$ is already known, then the existence of further solutions can be deduced from the general covariance of the equations". And then immediately after this sentence Einstein reproduced the Hole Argument.

After the Hole Argument Einstein wrote, "Having thus recognized that the useable theory of gravitation requires a necessary specialization of the coordinate system, we also see that the gravitational equations given by us are based upon special coordinate system".[168]

In section §5 Einstein concluded, "*the gravitational equations [(1)] are covariant with respect to all admissible transformations of the coordinate systems, i.e., with respect to all transformations between coordinate systems which satisfy the conditions [... (2)]*".[169] And the proof for this claim was that, since the conditions $B_\sigma = 0$, by which one restricted the coordinate systems, are direct consequence of the gravitational equations, therefore the covariance of the equations is far-reaching.[170]

## 4.3 Restricting Covariance is the Obvious

In March 1914, after presenting to him equations (1) and (2), Einstein wrote Besso,

"So I am going to live in Dahlem and have a room in Haber's institute [...]. At the moment I do not especially feel like working, for I had to struggle horribly to discover the above matter. The general theory of invariants appeared only as an impediment. The direct route proved to be the only feasible one. It is difficult to understand why I had to grope around so long before I found what was so obvious".[171]

I wish to thank Prof. John Stachel from the Center for Einstein Studies in Boston University for sitting with me for many hours discussing special and general relativity and their history. Almost every day, John came with notes on my draft manuscripts, directed me to books in his Einstein collection, and gave me copies of his papers on Einstein, which I read with great interest. I also wish to thank Prof. Alisa Bokulich, Director of the Boston University Center for History and Philosophy of Science, for her kind assistance while I was a guest of the Center. Finally I would like to thank her Assistant, Dimitri Constant, without whose advice and help I would not have been able to get along so well at BU and in Boston in general!

# Endnotes

"Let there be in our four dimensional manifold a portion L, in which a "material process" is not occurring, in which therefore the [the components of the stress-energy tensor] $\Theta_{\mu\nu}$ vanish. The $\Theta_{\mu\nu}$ given outside L, as requested by our assumptions, therefore also determines completely everywhere the [components of the metric tensor] $\gamma_{\mu\nu}$ inside L. We now imagine that, instead of the original coordinates, $x_\nu$, new coordinates $x'_\nu$ are introduced of the following type. Outside of L $x_\nu = x'_\nu$ everywhere; inside L, however, for at least a part of L and for at least one index $\nu$, $x_\nu \neq x'_\nu$. It is clear that by means of such a substitution it can be achieved that, at least for a part of L, $\gamma'_{\mu\nu} \neq \gamma_{\mu\nu}$. On the other hand, $\Theta'_{\mu\nu} = \Theta_{\mu\nu}$ everywhere, namely outside of L, because in this region $x_\nu = x'_\nu$, inside L, but because for this region, $\Theta_{\mu\nu} = 0 = \Theta'_{\mu\nu}$. It follows that in the case considered, if all substitutions are allowed as justified, namely, to the system of $\Theta_{\mu\nu}$ belongs more than one system $\gamma_{\mu\nu}$. So if – as has been the case in the paper – one maintains the requirement that the $\gamma_{\mu\nu}$ should be completely determined by the $\Theta_{\mu\nu}$, then one is forced to restrict the choice of reference system".

"*If the reference system is chosen totally arbitrarily, then the $g_{\mu\nu}$ can by no means be completely determined by the [stress-energy tensor] $T_{\sigma\nu}$. For imagine that the $T_{\sigma\nu}$ and $g_{\mu\nu}$ are given everywhere, and that all the $T_{\sigma\nu}$ vanish in a region of $\Phi$ of the four-dimensional space. I can now introduce a new* reference system, which agrees completely with the original outside of Φ, but is different from it inside Φ (without a violation of continuity). If one now refers everything to this new reference system, in which matter is represented by T'$_{\sigma v}$ and the gravitational field by g'$_{\mu v}$, then, even though we do have,

T'$_{\sigma v}$ = T$_{\sigma v}$

everywhere, the equations

g'$_{\mu v}$ = g$_{\mu v}$

are certainly not all satisfied inside Φ.[footnote 1] This proves the assertion.

If one wants to make it possible for the g$_{\mu v}$ (gravitational field) to be completely determined by the T$_{\sigma v}$ (matter), then this could only be achieved by restricting the choice of the reference system.

Footnote 1: The equations are thus to be understood, that each of the independent variables x'$_v$, on the left-hand side are to be given the same numerical values as the variables x$_v$ on the right-hand side".